# Partitioning Clustering Algorithm for Numerical and Categorical Data: a Review

Trupti M. Kodinariya[1]
Assistant Professor,
Department of Computer Engineering,
AITS- Rajkot – India
trupti.kodinariya@gmail.com

Dr. Prashant R. Makwana [2]
Director - Research
GRMECT Research Center
Rajkot – India
prmak1@hotmail.com

**ABSTRACT**

Clustering is widely used in different field such as biology, psychology, and economics. Most traditional clustering algorithms are limited to handling datasets that contain either numeric or categorical attributes. However, datasets with mixed types of attributes are common in real life data mining applications. In this paper, we review partitioning based algorithm such as K-prototype, Extension of K-prototype, K-histogram, Fuzzy approaches, genetic approaches, etc. These algorithm works on both numerical and categorical data. The approaches has been proposed to handle mixed data are based on four different perceptive: i) split data set into two part such that each part contain either numerical or categorical data, then apply separate clustering algorithm on each data set, finally combined the result of both clustering algorithm, ii) converting categorical attribute into numerical attribute and apply numerical attribute clustering algorithm; iii) discrimination of numerical attribute and apply categorical based clustering algorithm; iv) Conversion of the categorical attributes into binary ones and apply any numerical based clustering algorithm.

***Keywords:*** *Clustering, Cluster Ensemble method, Genetic clustering, FCM, K-histogram, K-prototypes.*

## 1. INTRODUCTION

Clustering is considered an important tool for data mining. The goal of data clustering is aimed at dividing the data set into several groups such that objects have a high degree of similarity to each other in the same group and have a high degree of dissimilarity to the ones in different groups [7]. Each formed group is called a cluster. Useful patterns may be extracted by analyzing each cluster. For example, grouping customers with similar characteristics based on their purchasing behaviors in transaction data may find their previously unknown patterns. The extracted information is helpful for decision making in marketing.

Various clustering applications [5, 9, 11, 14, 17] have emerged in diverse domains. Conventional clustering techniques have been focused on a single type of attributes, either numerical or categorical attributes of datasets. As a criterion function in clustering process, similarity measure has been used as one of the essential steps, i.e., in determining the candidate cluster number. The unique characteristics of categorical attributes are that the values of categorical attributes are not only discontinuous but also disordered while the values of numerical attributes are continuous in computing the distance between two values. Due to the difference of the characteristics between categorical and numerical attributes, similarity measures for categorical attributes or numerical attributes have focused on just their own characteristics, for example, entropy based similarity measure for categorical attributes and distance measure for numerical attributes.

As mixed attribute type datasets are common in real life, clustering techniques for mixed attribute type datasets is required in various informatics fields such as bio informatics, medical informatics, geo informatics, information retrieval, to name a few. These mixed attribute datasets provide challenges in clustering because there exist many attributes in both categorical and numerical forms so mixed attribute type should be considered together for more accurate and meaningful clustering.

However, conventional approaches are designed mainly for a single type attributes they are not appropriate for mixed attribute type datasets [8]. Typically, when people need to apply traditional distance-based clustering algorithms (ex., k-means [11]) to group these types of data, a numeric value will be assigned to each category in this attributes. Some categorical values, for example “low”, “medium” and “high”, can easily be transferred into numeric values. But if categorical attributes contain the values like “red”, “white” and “blue” … etc., it cannot be ordered naturally. How to assign numeric value to these kinds of categorical attributes will be a challenge work.

Recently, some approaches to clustering for mixed attribute have been introduced by converting categorical attribute values to numerical ones and applying traditional clustering algorithms with only numerical values [23]. However, those approaches have the possibility of distorting the results by losing characteristics of attributes [23]. Due to the fundamental differences in two data types, the conversion from categorical attribute values to numerical ones may not be perfect and not semantically meaningful. Many times, domain experts should be involved in the conversion process to provide semantic relations among different data types for a better conversion. However, this can be also subjective and incomplete. Thus, the loss of information in the conversion process incurs considerable inaccuracy in clustering.

In the paper, we explore different approaches of applying k-mean clustering on mixed data type i.e. both numerical and categorical data.

The paper is organized as follows. A generic version of K-Means is described in Section 2. Section 3 contains a review of different approaches where K-Means applied on mixed data type. Section 4 concludes the paper.

## 2. TRADITIONAL PARTITIONING CLUSTERINGAL GORITHM FOR HANDLING NUMERICAL AND CATEGORICAL DATA

Several clustering algorithm has been proposed for only numerical and categorical data. The k-means and k-mode is well-known clustering algorithm apply on numerical and categorical data respectively. (K-mode can also apply on numerical data)

### 2.1 K-means Clustering

K-Means [11] is one of the simplest unsupervised learning algorithms that solve the well known clustering problem. The procedure follows a simple and easy way to classify a given data set through a certain number of clusters (assume k clusters) fixed a priori.

The main idea is to define k centroids, one for each cluster. These centroids should be placed in a cunning way because of different location causes different result. So, the better choice is to place them as much as possible far away from each other. The next step is to take each point belonging to a given data set and associate it to the nearest centroid. When no point is pending, the first step is completed and an early group is done. At this point, it is need to re-calculate k new centroids as centers of the clusters resulting from the previous step. After these k new centroids, a new binding has to be done between the same data points and the nearest new centroid. A loop has been generated. As a result of this loop it may notice that the k centroids change their location step by step until no more changes are done. In other words centroids do not move any more.

Finally, this algorithm aims at minimizing an objective function, in this case a squared error function. The objective function

$$W(S, C) = \sum_{k=1}^{K} \sum_{i \in S_k} \| y_i - c_k \|^2$$

where S is a K-cluster partition of the entity set represented by vectors $y_i$ ($i \in I$) in the M-dimensional feature space, consisting of non-empty non-overlapping clusters $S_k$, each with a centroid $c_k$ (k=1,2,…K).

The algorithm is composed of the following steps:

- Place k points into the space represented by the objects that are being clustered. These points represent initial group centroids
- Assign each object to the group that has the closest centroid
- When all objects have been assigned, recalculate the positions of the k centroids
- Repeat Step 2 and 3 until the centroids no longer move

This produces a separation of the objects into groups from which the metric to be minimized can be calculated. Traditional K-means algorithm is applied on only numerical data and it generate local optimal solution

### 2.2 K-mode Clustering

The k-modes algorithm [27] is an extension to k-means algorithm to cluster categorical data by removing the drawback that was put by k-means by using a simple matching dissimilarity measure or the hamming distance for categorical data objects and replacing means of clusters by their modes. This algorithm has made three major modifications to the kmeans: firstly it uses a different similarity or rather dissimilarity measure known as chi- square distance as mentioned below, it has replaced k-means to k modes and lastly uses a frequency based method to update modes.

The dissimilarity measure of k-modes algorithm follows:

Let A, B be two categorical objects described by n categorical attributes. The dissimilarity measure between these two objects A and B can be thence be defined by the all and total mismatches of the corresponding attribute categories of the two objects. The smaller the number of mismatches is, the more similar the two objects. Formally,

$$d(A,B) = \sum_{i=1}^{m} \delta(a_i, b_i)$$

$$\text{where } \delta(a_i, b_i) = \begin{cases} 0 (a_i = b_i) \\ 1 (a_i \neq b_i) \end{cases}$$

d(A,B) here gives more or less equal importance to each category of an attribute. The frequency of categories if taken into account can then lead to a dissimilarity measure as defined below:

$$d_{x^2}(X,Y) = \sum_{j=1}^{m} \frac{(n_{xj} + n_{yj})}{n_{xj} n_{yj}} \delta(x_j, y_j)$$

where $n_{xj}$ , $n_{yj}$ are the numbers of objects in the data set that have categories $x_j$ and $y_j$ for attribute j. Because $d_x^2(X, Y)$ is similar to the chi-square distance. The dissimilarity measure here uses and prioritizes rare categories than to frequent ones.

K-modes algorithm [14] has its own set of drawbacks because of its instability due to non-uniqueness of the modes i.e., the results of the clusters depend largely and strongly on the selection of modes during the clustering process.

## 3. Different approaches of handling mixed data type in Partitioning Clustering

K-prototype [25] is one of the first extended methods proposed in the literature. It is a combination of K-means [11] and K-mode [27] methods. The distance measure used with K-prototype is the sum of Euclidean distance, which is used to compute the distance between numerical attributes, and the simple matching measure used to dealt with categorical attributes where the distance between two objects is computed as the number of different attributes values between these objects. To build classes, K-prototype follows the same steps as K-means and K-mode.

The a cost function for clustering mixed data sets with n data objects and m attributes ($m_r$ numeric attributes, $m_c$ categorical attributes, $m = m_r + m_c$) as

$$\zeta = \sum_{i=1}^{n} \vartheta(d_i, C_j),$$

where $\vartheta(d_i, C_j)$ is the distance of a data object di from the closest cluster center $C_j$. $\vartheta(d_i, C_j)$ is defined as

$$\vartheta(d_i, C_j) = \sum_{t=1}^{m_r} (d_{it}^r - C_{jt}^r)^2 + \gamma_j \sum_{t=1}^{m_c} \delta(d_{it}^c, C_{jt}^c),$$

$d^r_{it}$ are values of numeric attributes and dcit are values of categorical attributes for data object di. Here $C_j = (C_{j1}, C_{j2}, . . . , C_{jm})$ represents the cluster center for cluster j. $C^c_{jt}$ represents mean of numeric attribute t and cluster j. For categorical attributes, $\delta(p,q) = 0$ for $p = q$ and $\delta(p,q) = 1$ for $p \neq q$. $\gamma_j$ is a weight for categorical attributes for cluster j. Cost function $\zeta$ is minimized for clustering mixed data sets.

The main shortcomings of k-prototype may fall into followings:

- Binary distance is employed for categorical value. If object pairs with the same categorical value, the distance between them is zero; otherwise it will be one. However, it will not properly show the real situation, since categorical values may have some degree of difference. For example, the difference between "high" and "low" shall not equal to the one between "high" and "medium".
- For categorical attributes, the cluster center is represented by the mode of the cluster rather than the mean. While this allays the problem of finding the mean for categorical values, there is information loss since the true representation of the cluster is not obtained. Only one attribute value represents the cluster, even though there may be close seconds or thirds.
- In cost function weight of all numeric attributes is taken to be 1. The weight of categorical attributes is a user-defined parameter $\gamma_j$. However, in a real data set all numeric attributes may not have the same effect on clustering. Incorrect user-given values of $\gamma_j$ may also lead to inaccurate clustering.

Amir Ahmad and Lipika Dey [1] propose algorithm which overcomes the problem of cost function used Huang [25]. Rather than the weight of categorical attribute is taken form user, it is automated from data set itself. The distance computation schemes for handling numeric and categorical values have been designed to take care of the shortcomings discussed above. In our scheme, $\delta$ (p,q) which denotes the distance between a pair of distinct values p and q of an attribute, is computed as a function of their co occurrence with other attribute values. The significance of an attribute, which determines the contribution of an attribute towards clustering (given by the weight in the distance function), is computed in terms of $\delta$ (p,q). Thus, the weighing values are extracted from the attribute value distributions within the data and need not be user defined. The significance of a numeric attribute is also computed similar to a categorical attribute and all numeric attributes do not contribute equally towards clustering.

Here, $\vartheta(d_i,C_j)$ is defined as

$$\vartheta(d_i, C_j) = \sum_{t=1}^{m_r} (w_t(d_{it}^r - C_{jt}^r))^2 + \sum_{t=1}^{m_c} \Omega(d_{it}^c, C_{jt}^c)^2$$

Where $\sum_{t=1}^{m_r} (w_t(d_{it}^r - C_{jt}^r))^2$ denotes the distance of object $d_i$ from its closest cluster center $C_j$, for numeric attributes only, $w_t$ denotes the significance of the $t^{th}$ numeric attribute, which is to be computed from the data set,

$\sum_{t=1}^{m_c} \Omega(d_{it}^c, C_{jt}^c)^2$ denotes the distance between data object $d_i$ and its closest cluster center $C_j$ in terms of categorical attributes only.

The k-histograms algorithm [26] extends the k-mean algorithm to categorical data by replacing the means of clusters with histogram. It dynamically updates histograms in the clustering process and is found to be very high in accuracy. In general, this algorithm is very similar to the k-modes algorithm except that it uses the histogram data structure to describe a categorical data cluster instead of mode.

The dissimilarity measure is defined as follow

Let $A_1$, $A_2$... $A_m$ be a set of categorical attribute with domain $D_1$, $D_2$ …Dm respectively. Let the dataset D= {$X_1$, $X_2$, .. $X_n$} be a set of objects described by m categorical attributes. The valve set $V_i$ of $A_i$ is the set of valves of $A_i$ that are present in D. For each $v \in V_i$, the frequency f(v) denoted as fv, is number of objects $O \in X$ with $O.A_i = v$. suppose the number of distinct attribute valve of Ai is Pi, we define the histogram of Ai, as the set of pairs: hi = {($v_1$,$f_1$), ($v_2$,$f_2$),…($v_p$,$f_p$)}. The Histogram of dataset is defined as H= {$h_1$ ,$h_2$,…., $h_m$}

The dissimilarity measure is defined here in terms of a histogram H which is compact representation of dataset D and an, object Y.

$$d(H,Y) = \frac{\sum_{j=1}^{m} \psi(h_j, y_j)}{n}$$

Where

$$\psi(h_j, y_j) = \sum_{l=1}^{p_j} f_l * (1 - \delta(v_l, y_j))$$

The K-histogram algorithm is works as follow:

1. Select k initial histograms, one for each cluster.
2. Allocate an object to the cluster whose histogram is the nearest to it according to dissimilarity measure. Update the histogram of the cluster after each allocation.
3. After all objects have been allocated to clusters, retest the dissimilarity of objects against the current histograms. If an object is found such that its nearest histogram belongs to another cluster rather than its current one, reallocate the object to that cluster and update the histograms of both clusters.
4. Repeat 3 until no object has changed clusters after a full cycle test of the whole data set.

Like k-means algorithm, the K-histogram algorithm also produces locally optimal solutions that are dependent in initial histograms and the order of objects in the data set.

Another extension of K-mode is K-representative algorithms [18] which overcome the problem of non-uniqueness of the modes by introducing a new notion of "cluster centers" called representatives for categorical objects. Arithmetic operations are completely absent in the initialization and setting of categorical objects, it applies the notion of fuzzy logic in defining representatives instead of means for clusters. With this theory, it can formulate the clustering problem of categorical objects as a partitioning problem in the way similar to k-means clustering. The dissimilarity measure of this algorithm is as follows.

The dissimilarity between a categorical object and the representative of a cluster is defined based on simple matching as follows.

Let C = {$X_1, \ldots, X_p$} be a cluster of categorical objects, with $X_i = (x_{i,1}, \ldots, x_{i,m})$, $1 \leq i \leq p$, denote by $D_j$ the set formed from categorical attributes ($x_{1,j}, \ldots x_{p,j}$.) The representative of C is defined by $Q = (q_1, \ldots, q_m)$, with $q_j = \{(c_j, f_{cj}) \mid c_j \in D_j\}$, where $f_{cj}$ is the relative frequency of category cj within C, i.e., $f_{cj} = n_{cj}/p$, where $n_{cj}$ is the number of objects in C having category cj at attribute $A_j$. Formally, each $q_j$ can be seen as a fuzzy set on $D_j$ with membership grades of elements to be defined by their relative frequencies within the cluster.

The dissimilarity d(X, Q) is mainly dependent on the relative frequencies of categorical values within the cluster and simple matching between categorical values. Here it is important to note that the simple matching dissimilarity measure between categorical objects can be considered as a categorical counterpart of the squared Euclidean distance measure. K-Representative defines the dissimilarity between object X and representative Q by

$$\begin{aligned} d(X,Q) &= \sum_{j=1}^{m} \sum_{c_j \in D_j} f_{c_j} \delta(x_j, c_j) \\ &= \sum_{j=1}^{m} \sum_{c_j \in D_j, c_j \neq x_j} f_{c_j} \\ &= \sum_{j=1}^{m} (1 - f_{x_j}), \end{aligned}$$

where $f_{xj}$ is the relative frequency of category xj within C.

Cluster Ensemble is a method to combine several runs of different clustering algorithm to get a common partition of the original dataset. In the cluster ensemble approach numeric data are handled separately and categorical data are handled separately. Then both the results are then treated in a categorical manner. There are two main issues in designing cluster ensembles: (1) the design of the individual “clusterers” so that they form potentially an accurate ensemble, and (2) the way the outputs of the clusterers are combined to obtain the final partition, called the consensus function. In [2], the authors formally defined the cluster ensemble problem as an optimization problem and propose combiners for solving it based on a hyper-graph model. A multi-clustering fusion method is presented in [3]. In their method, the results of several independent runs of the same clustering algorithm are appropriately combined to obtain a partition of the data that is not affected by initialization and overcomes the instabilities of clustering methods. After that, the fusion procedure starts with the clusters produced by the combining part and finds the optimal number of clusters according to some predefined criteria. The authors in [20] proposed a sequential combination method to improve the clustering performance. First, their algorithm uses the global criteria based clustering to produce an initial result, then use the local criteria based information to improve the initial result with a probabilistic relaxation algorithm or linear additive model.

In [6], Ralambondrainy presented modified k-means (known as c-mean) to cluster categorical data by converting multiple categorical attributes into binary attributes, each using one for presence of a category and zero for absence of it, and then treats these binary attributes as numeric ones in the k-means algorithm. This needs to handle a large number of binary attributes when data sets have attributes with many categories increasing both computational and storage cost. The other drawback is that the cluster means given by real values between zero and one do not indicate the characteristics of the clusters. The conventional (hard) clustering methods restrict each point of the data set to exactly one cluster. In [15], proposed fuzzy sets that produced the idea of partial membership of belonging described by a membership function, fuzzy clustering as been widely studied and applied in a variety of substantive areas. In the literature on fuzzy clustering, the fuzzy c-mean (FCM) clustering algorithms are the best-known methods [10, 13]. In the original formulation of the FCM algorithm the dissimilarity functional dij is taken as the Euclidean distance of the $j^{th}$ observed data point, $x_j$, from the centroid of the $i^{th}$ cluster, $\mu_i$, i.e.,

$$d_{ij} = ||\boldsymbol{x}_j - \boldsymbol{\mu}_i||_2^2$$

This selection implies, in essence, the assumption of data organized into spherical clusters. However, this might not be the case in many real-life applications. Several approaches have been proposed to modify FCM. In [19], proposed approach for fuzzy clustering of data with mixed numeric and categorical attributes. The approach is based on the postulation of an appropriate prior assumption regarding the distribution of the sought clusters, and the subsequent derivation of a suitable methodology for the effective introduction of these assumptions into the fuzzy clustering procedure. Our algorithm is formulated using a regularized by KL information fuzzy objective function, providing significant intuitive advantages over the original FCM algorithm objective function

In [12], proposed a new clustering method called genetic k-means algorithm (GKA) [12], which hybridizes a genetic algorithm with the k-means algorithm. This hybrid approach combines the robust nature of the genetic algorithm with the high performance of the k-means algorithm. As a result, GKA will always converge to the global optimum faster than other genetic algorithms. FGKA (Fast Genetic K-means Cluster Technique) was proposed by Lu et al. [22]. It has many enhancements over GKA that includes evaluation with Total Within-Cluster Variation (TWCA). The enhancements include mutation operator simplification and avoiding illegal string elimination overhead. Moreover FGKA is faster than other algorithms such as GKA. Nevertheless, the FGKA algorithm has a significant disadvantage. When the mutation probability is small, it gives problems besides making the algorithm more expensive. In order to overcome this issue Lu et al. proposed IGKA (Incremental Genetic Kmeans Algorithm) which exceeds the FGKA in performance. The incremental nature of the IGKA provides such performance benefits. When compared with FGA, IGKA performs well. From this perspective the a hybrid algorithm is proposed that is the combination of FGKA and IGKA and performs exceptionally well when the mutation probability is more. Many GA based algorithms are based on K-means and they work only for numeric data. Another generic algorithm known as GKMODE works only for discrete data. In [4], proposed genetic k-mean clustering algorithm for mixed numeric and categorical data. In this approach modified cost function used which handle mixed data and Proportional selection is used for the selection operator in which, the population of the next generation is determined by Z independent random experiments. Modified representation for the cluster centre is used. This representation can capture cluster characteristics very effectively because it contains the distribution of all categorical values in cluster. Also in this paper, additional some features such as efficient calculation of TWCVs, avoiding illegal string elimination overhead, and the simplification of the mutation operator. The initialization phase and the three operators are redefined to achieve these improvements.

In summary, In order to handle mixed numeric and categorical data, the strategies that have been employed are classified as follows:

A. Cluster Ensemble approach is to group each kind of variable together, performing a separate cluster analysis for each variable type and then combining results of each of them [16, 21]. This is feasible if these analyses derive compatible results. However, in real applications, it is unlikely that a separate cluster analysis per variable type will generate compatible results.
B. The conversion of each categorical attribute into a numerical attribute. The clustering methods based on numerical attributes can then be used. However, it is difficult sometimes to code a categorical attribute without understanding the meaning of this attribute [1]. For example, the conversion of an attribute color is more difficult than the conversion of an attribute month.
C. The discretization of numerical attributes so that each numerical attribute is converted into a categorical one by using a discretization method. Then the objects can be clustered by using one of the clustering methods based on categorical attributes. DSqueezer (Discretizing before Using Squeezer) [24] method follows this idea. The principal limit of this type of approach is the loss of information generated by discretizing.
D. Conversion of the categorical attributes into binary ones: each categorical attribute Ai with d distinct values is converted into d binary attributes where each attribute takes value 1 if the object is described by this attribute, 0 if not. The number of attribute need to handle is increase.

# 4. CONCLUSIONS

Real world databases always contain numeric and discrete values. The numeric values are generally used in mathematic operations and categorical values can't be used. This is the reason many clustering algorithms fail in working with large datasets that contain attributes of mixed type. The popular Euclidian Distance (ED) and K-means clustering algorithms in their original form fail to work with mixed content. K-prototype is first extended version of k-mean and k-mode algorithm which works on mixed data. The shortcoming of k-prototype is: i) it require weight parameter whose valve provided by end use; ii) binary distance measure is used for categorical attribute. K-representative algorithm overcomes this problem. K-histogram is suitable for image processing application. The conventional K-Modes algorithm is capable of efficiently clustering categorical data; however, it uses hard centroids for categorical attributes and a simple distance measure to classify boundary data. To address these shortcomings of the K-Modes algorithm, the existing fuzzy clustering algorithm is modified so that it uses fuzzy centroids for clustering categorical data (FCM). FCM can work on only categorical attribute. In [3] proposed fuzzy based clustering algorithm that works on mixed data. The local optima problem with traditional partitioning clustering method can overcome by hybridizing with genetic algorithm. Dharmendra K Roy & Lokesh K Sharma proposed a genetic clustering algorithm works on mixed data.